\documentclass[manuscript,nonacm]{acmart}

\AtBeginDocument{%
  \providecommand\BibTeX{{%
    \normalfont B\kern-0.5em{\scshape i\kern-0.25em b}\kern-0.8em\TeX}}}

\copyrightyear{2025}
\acmYear{2025}
\setcopyright{cc}
\setcctype{by}
\acmConference[CHI '25]{CHI Conference on Human Factors in Computing Systems}{April 26-May 1, 2025}{Yokohama, Japan}
\acmBooktitle{CHI Conference on Human Factors in Computing Systems (CHI '25), April 26-May 1, 2025, Yokohama, Japan}\acmDOI{10.1145/3706598.3713512} \acmISBN{979-8-4007-1394-1/25/04}

\begin{document}

\title[Ear-EEG for Flow in Knowledge Work]{Exploring Flow in Real-World Knowledge Work Using Discreet cEEGrid Sensors}

\author{Michael T. Knierim}
\email{michael.knierim@kit.edu}
\orcid{0000-0001-7148-5138}
\affiliation{%
  \institution{Karlsruhe Institute of Technology}
  \country{Germany}
}

\author{Fabio Stano}
\email{fabio.stano@kit.edu}
\orcid{0009-0007-7948-7104}
\affiliation{%
  \institution{Karlsruhe Institute of Technology}
  \country{Germany}
}

\author{Fabio Kurz}
\email{uletf@student.kit.edu}
\orcid{0009-0001-5517-1013}
\affiliation{%
  \institution{Karlsruhe Institute of Technology}
  \country{Germany}
}

\author{Antonius Heusch}
\email{antonius.heusch@student.kit.edu}
\orcid{0009-0009-0248-3257}
\affiliation{%
  \institution{Karlsruhe Institute of Technology}
  \country{Germany}
}

\author{Max L. Wilson}
\email{Max.Wilson@nottingham.ac.uk}
\orcid{0000-0002-3515-6633}
\affiliation{%
  \institution{University of Nottingham}
  \country{United Kingdom}
}

\renewcommand{\shortauthors}{Knierim et al.}

\newcommand{\revision}[1]{\textcolor{black}{#1}}

\begin{abstract}
  Flow, a state of deep task engagement, is associated with optimal experience and well-being, making its detection a prolific HCI research focus. While physiological sensors show promise for flow detection, most studies are lab-based. Furthermore, brain sensing during natural work remains unexplored due to the intrusive nature of traditional EEG setups. This study addresses this gap by using wearable, around-the-ear EEG sensors to observe flow during natural knowledge work, measuring EEG throughout an entire day. In a semi-controlled field experiment, participants engaged in academic writing or programming, with their natural flow experiences compared to those from a classic lab paradigm. Our results show that natural work tasks elicit more intense flow than artificial tasks, albeit with smaller experience contrasts. EEG results show a well-known quadratic relationship between theta power and flow across tasks, and a novel quadratic relationship between beta asymmetry and flow during complex, real-world tasks.
\end{abstract}




\maketitle

\section{Introduction}
Flow, linked to peak performance and well-being, is widely studied in HCI to enhance user experiences across technology interactions \cite{Klarkowski2015, Brown2023, Labonte2016, Wang2022, Wang2014, Oliveira2019, Johnson2015, Berger2023}. While traditionally examined in leisure contexts, recent work highlights its frequent occurrence in professional settings, where fostering flow boosts productivity, mastery, and competence while reducing burnout \cite{Brown2023, Cowley2022, Wang2023, Pastushenko2020, Yotsidi2018, Afergan2014, Zueger2017}. \revision{For instance, someone coding or crafting a story might become deeply absorbed, losing track of time, their surroundings, and self-doubt, while making efficient progress and feeling profoundly satisfied with their work}.
Therefore, to foster flow in work situations, technological detection methods have been focused, especially in knowledge work where information technology is ubiquitous and can provide unobtrusive and automatic data through sensors. Such continuous data can enable tools like adaptive dashboards that maintain optimal demands \cite{Afergan2014}, and hence, flow, warning lights \cite{Zueger2017} that prevent flow interruptions, or feedback tools that aid flow self-regulation \cite{Adam2024, Yang2019}. \\

To that end, research is making detection progress, for example using programmers' behavior or log data \cite{Cowley2022, Brown2023, Zueger2017}, but also leveraging body and brain sensing \cite{Rissler2020, Rissler2018, Tozman2015, Bian2016, Berta2013, Labonte2016}. Physiological sensing provides the unique opportunity of task-independent flow indicators, bypassing the need for additional behavioral data. However, despite improvements, physiological sensing methods largely remain confined to lab environments and controlled tasks. 
This is problematic because flow experiences in lab settings may be inauthentic \cite{Hommel2010, Harris2017}. Controlled tasks and environments raise questions about whether \textit{real} flow is experienced, as the artificial situations may impose limitations on attentional and motivational processes required for the holistic task focus that is required for flow \cite{Hommel2010}. Additionally, lab confinement limits our understanding of flow detection in everyday scenarios, where confounding factors play a significant role. \\

Some studies have explored the physiology of flow in work settings using heart rate (ECG/PPG) and skin conductance (EDA/GSR) sensors \cite{Rissler2020, Graft2023}. However, a key gap remains in brain data collection during natural knowledge work. Brain data offers high temporal resolution and rich insights into mental processes, and has proven valuable in lab-based flow studies \cite{Berta2013, Ewing2016, Barros2018, Wolf2014, Chanel2011, Afergan2014, Leger2014}. Yet, both researchers and participants are hesitant to use brain sensing methods in everyday life due to the cumbersome setup of devices like EEG or fNIRS, which require lengthy preparations (e.g., electrode fitting, gel application) and are socially obtrusive, deterring wearers in real-world settings \cite{Knierim2023, lee2020stress, kosmyna2019attentivu}.
In this research, we aim to bridge this gap by utilizing recent advancements in wearable EEG technology, specifically low-cost, discreet designs that could eventually integrate into everyday wearables like glasses or headphones. We focus on gelled, around-the-ear EEG sensors (open-cEEGrids) to explore their potential for whole-day flow data collection in real-life settings.
Thereby, we set out to answer the following research questions:
\begin{itemize}
    \item RQ1: How do flow experiences emerge during various forms of natural knowledge work?
    \item RQ2: How do natural flow experiences compare to flow experiences in a classic, controlled lab task?
    \item RQ3: How can around-the-ear EEG inform flow monitoring in natural knowledge work?
\end{itemize}

To address these questions, twenty-one participants, set up with open-cEEGrid sensors, completed a single-day, multi-session observation in a home-office (i.e., field) setting. Thereby, almost 200 hours of EEG data were collected \revision{(over 100 hours during observed tasks)}. Each participant engaged in four 1-hour sessions working on their own knowledge project — either an academic thesis or a software engineering project. Their tasks were periodically interrupted to gather reports about their flow experiences. To provide a reference for these natural flow experiences, participants also completed two sessions of a classic difficulty-manipulated mental arithmetic task for flow induction.
Our results provide the following insights and contributions:

\begin{itemize}
    \item We find that both academic thesis and software engineering projects elicit intense flow, and we provide a detailed analysis of the specific aspects of these natural knowledge work tasks' flow variations. As these variations are rather small, our findings inform future studies aiming to elicit stronger flow contrasts in natural knowledge work days.
    \item We present the first direct comparison of flow intensities between natural and controlled tasks. We find that natural tasks elicited more intense flow, while the controlled tasks were more effective in generating low-flow states, providing greater flow contrasts.
    \item Ear-EEG results show a \revision{convex quadratic} relationship between theta frequency band power and flow reports across both natural and controlled tasks. This finding is significant as it confirms that this well-established lab observation can also be replicated in more open and natural field settings.
    \item We also identify a significant \revision{convex} quadratic relationship between beta frequency band asymmetry and flow, observed only in natural knowledge work. We discuss this as a potential indicator of \revision{attentional or} cognitive flexibility during complex, real-world tasks.
\end{itemize}

Altogether, these results contribute to bridging the gap in flow field studies using brain sensing, and offer a way to advance context-free, automatic flow detection further. By introducing this initial approach, we hope to inspire more research using discreet ear-EEG as a method for detecting flow in everyday life.

\section{Background \& Related Work}
\subsection{Flow Experiences}
Flow is described as an experience of complete task absorption characterized by effortless, fluent action \cite{Csikszentmihalyi1975, Bruya2010}. This state is often linked with peak performances and heightened well-being \cite{Csikszentmihalyi1975, Bruya2010, Oliveira2019, Yotsidi2018, Afergan2014}.
Flow can be observed in almost any task that requires active engagement, and that primarily fulfills the three essential pre-conditions: (1) balance of a task's challenge and a person's skill, (2) a clear goal, and (3) unambiguous feedback about whether that goal is being reached \cite{Csikszentmihalyi1975}.
This ubiquitous potential for flow emergence has led to its extensive study in scenarios ranging from work \cite{Brown2023, Cowley2022, Afergan2014} to education \cite{Wang2023, Pastushenko2020}, and from games \cite{Klarkowski2016, Labonte2016, Klarkowski2015, Tozman2015, Johnson2015} to interactions with digital technologies \cite{Bian2016, Berger2023}. 
In the HCI domain, as it is associated with great user experiences, flow has been prolifically studied for example in single-player \cite{Klarkowski2015, Tozman2015, Ewing2016} and multi-player gaming \cite{Labonte2016, Johnson2015}, virtual reality experiences \cite{Bian2016}, live-streaming \cite{Wang2022}, music-streaming services \cite{Wang2014}, educational systems \cite{Oliveira2019}, and adaptive work software \cite{Afergan2014}. \\

Interestingly, while flow is commonly studied in leisure settings, research finds that it is particularly often experienced at work because many work tasks, particularly those involving complex problem-solving or creative activities, inherently fulfill the conditions for flow \cite{Engeser2016, Quinn2005}. 
Knowledge work, in particular, is a rich domain for studying flow due to its inherent complexity, cognitive demands and need for sustained focus \cite{Engeser2016, Quinn2005}. Previous research has focused on specific types of knowledge work, such as software engineering \cite{Brown2023, Cowley2022}, and academic research and writing \cite{Knierim2018, Graft2023}. Software engineers, for instance, are often found to experience flow due to the particularly fast and clear cycles of task feedback upon code execution \cite{Brown2023, Cowley2022}. Similarly, academic work, with its rigorous demands for focus and intellectual engagement, provides a fertile ground for flow emergence \cite{Quinn2005, Knierim2018, Graft2023}. 
%

\subsection{Flow Research Paradigms}
Flow research methods generally fall into two main approaches: (1) natural flow observation using the experience sampling method (ESM) and (2) controlled flow induction through manipulating the above-mentioned flow pre-conditions.
In natural flow observation, participants are typically equipped with smartphones that prompt them to report their flow experiences throughout their daily lives, often over several days or weeks \cite{Csikszentmihalyi2003, Rissler2020, Graft2023, Zueger2017, Zueger2018}. This method aims to \textit{capture} flow as it naturally occurs \cite{Csikszentmihalyi2003}. While this approach offers high internal and ecological validity, it is limited by long study durations paired with relatively low temporal data resolution.
In contrast, controlled flow induction typically involves presenting participants with difficulty-manipulated (DM) tasks, designed to elicit varying mental states such as boredom, flow, and overload \cite{Keller2016, Klarkowski2015, Katahira2018}. This method, central to experimental flow research, focuses on manipulating the challenge-skill balance, which is considered a critical pre-condition for flow. \\

Especially, in physiological flow research, the difficulty- manipulation paradigm has been crucial for eliciting flow within controlled laboratory settings, which are necessary for accurate measurement instrumentation. A key advantage of this approach is its proven ability to create contrasts in flow experiences: low flow in tasks that are either too easy or too difficult, and higher flow in tasks with balanced difficulty \cite{Keller2016, Klarkowski2015}. However, this paradigm has faced significant criticism. Neuroscientists argue that the highly controlled and often artificial nature of difficulty-manipulated tasks may hinder the emergence of genuine flow experiences, which typically require high intrinsic motivation and task familiarity to overcome attentional barriers that gate flow \cite{Hommel2010, Harris2017}.
While these differences are discussed in the flow literature, there haven't yet been any studies that compare the two paradigms directly.
More critically, this dichotomic situation highlights that current knowledge of flow physiology is predominantly based on controlled lab settings and tasks, due to the absence of viable measurement options for studying flow physiology in everyday life.

\subsection{Flow Measures \& Physiology}
Self-reports remain the predominant measurement approach for flow studies, while objective measures are still being developed \cite{Oliveira2019, Bian2016, Klarkowski2015, Peifer2021}. While these developments have proven complex, they remain fueled by the two main limitations of self-reports: (1) the mere act of reporting often disrupts the flow state, and (2) these interruptions can only occur occasionally, giving a highly fragmented picture of the temporal dynamics of flow. These limitations highlight the ever present need for automated and unobtrusive flow measurement methodologies. 
In response, researchers have, for example, formulated metrics using behavioral log data to classify focused work states in software engineers \cite{Brown2023, Cowley2022}. However, the broader quest for context-free flow detection has notably led to a rising interest in physiological flow detection in recent years \cite{Harris2017, Peifer2021}. These approaches can be broadly distinguished in studies focusing on peripheral nervous system (PNS) measures (e.g., heart rate, skin conductance, breathing), or central nervous system measures (CNS), that is brain activity (cerebral electricity or blood flow). \\

On the PNS side, flow is most often characterized by an inverted-U-shaped relationship with indicators of physiological arousal \cite{Tozman2015, Bian2016, Peifer2021}. 
Therefore, flow is considered to require an optimal balance between energy mobilization and energy recovery (not too much physiological excitement and not too little \cite{Peifer2021}). 
Based on such PNS measurements, researchers have found promising accuracies for machine-learning based flow-classification using heart-rate (ECG) data in an invoice-matching task \cite{Rissler2018}, ECG coupled with face, voice, and body expressions in a learning context \cite{Wang2023}, or pulse (BVP) and sweat gland (EDA) activity during a Tetris game \cite{Maier2019}. 
Notable recent studies have also started to combine common wearable sensors like heart rate chest belts or smart wristwatches for in-field bio-signal measurements for flow detection, finding similar classifier performances as in the lab settings \cite{Rissler2020, Graft2023}. \\

On the CNS side, flow researchers have leveraged measurements like fMRI \cite{Ulrich2014} or fNIRS \cite{Barros2018, Afergan2014} to study blood flow dynamics in the brain or, most prominently, EEG \cite{Ewing2016, Labonte2016, Nacke2008, Leger2014, Katahira2018} to observe the brain's electrical activity during flow. 
We herein focus on the related work on EEG measures, as it currently represents the best candidate for flow research in natural knowledge work due to its information richness, established methodology, and its increasing efficacy in mobile settings \cite{niso2023wireless}. Recent reviews on wireless EEG emphasize the established nature of wearable EEG, with the field witnessing considerable progression in miniaturized components, seamless integration, and improved signal quality \cite{niso2023wireless}.
Importantly, in an effort to hide the usually clunky and complicated brain scanning gear, researchers have developed transparent EEG solutions like the AttentivU – smart glasses with sensors around the nose and ear \cite{kosmyna2019attentivu}, in-ear EEG systems \cite{lee2020stress}, around-the-ear EEG electrodes \cite{Debener2015, Wascher2019, Hoelle2021, Knierim2021BCI}, and headphone EEG systems \cite{An2021, Cherep2022, kartali2019real, Knierim2023}. \\

\revision{EEG flow research has explored various methods to link flow with neural activity \cite{Alameda2022}, focusing primarily on EEG markers of attentional focus, mental workload or engagement. 
Some studies used auditory event-related potentials (ERPs) - a tone frequently playing in the background - to assess how brain processing of background tones varies with task absorption \cite{Bombeke2019, Nunez2019}, and found reduced ERP amplitudes during higher task absorption (i.e., flow), indicating decreased attention to irrelevant stimuli during flow.
In contrast, most flow EEG work has} analyzed neural oscillations, particularly frequency band power changes across the scalp. A common finding is the elevation of theta band power over frontal areas, linked to mental workload and engagement in tasks like difficulty-manipulated games and mental arithmetic \cite{Nacke2008, Ewing2016, Fairclough2013, Katahira2018}.
Similar patterns have been observed over temporal cortex sites too, most likely caused by volume conduction as the electrical activity from frontal brain regions spreads across the scalp \cite{Berta2013, Chanel2011}. 
The theta-flow relationship often follows an inverted U-shape, with flow decreasing as tasks become too difficult and individuals disengage \cite{Nacke2008, Chanel2011, Fairclough2013, Ewing2016}. \revision{However, some studies also report minimal differences in theta power between moderate and high task difficulties \cite{Katahira2018, Berta2013}.}
\revision{Generally, theta oscillations are thought to coordinate neural processes for sustained attention and task-relevant information integration. During flow, moderate fronto-central theta activity could thus reflect necessary attention for challenging tasks, focusing exclusively on task-relevant information \cite{Alameda2022, Harris2017}. This efficient state avoids task-unnecessary processes, such as self-monitoring in overly demanding tasks, which may also explain the ease of processing experienced during flow \cite{Harris2017}}. \\

Studies combining both PNS and EEG measures consistently show that EEG features, particularly frequency bands, contribute more strongly to machine learning classification of flow, likely due to their higher specificity in capturing cognitive processes compared to PNS measures, which reflect broader physiological arousal \cite{Chanel2011, Berta2013, Leger2014}.
However, it is important to note that these findings are primarily based on difficulty-manipulated tasks, which may present a paradigm-specific bias.
A second emerging observation in flow EEG research are hemispheric asymmetries during repeated, non-manipulated tasks \cite{Wolf2014, DeKock2014, Kramer2007, Labonte2016}. For example, \cite{Labonte2016} found heightened right frontal alpha activation during flow in a collaborative video game, while \cite{Wolf2014} observed similar alpha asymmetry in table tennis players anticipating a serve. \cite{Labonte2016} attributed this to greater approach-motivation, whereas \cite{Wolf2014} linked it to reduced verbal-analytic interference in motor tasks. Reduced theta asymmetry in repeated game trials was also noted \cite{Kramer2007} and \cite{DeKock2014}. These asymmetries seem to emerge more frequently in tasks requiring experience and mastery, when comparing novices to experts.
However, all the CNS research mentioned above was conducted in lab settings, mostly using traditional EEG caps, with no studies exploring flow in everyday life using wearable EEG. In this work, we leverage ear-EEG advancements to extend flow research into such real-world settings.

\section{Method}
\subsection{Study Design \& Tasks}
To explore the potential of monitoring flow experiences in natural knowledge work using discreet, around-the-ear EEG systems, we designed a single-day field study that combined both open, natural work tasks and classic, difficulty-manipulated tasks. The latter served as a reference to the flow induction paradigm commonly used in lab-based flow research. We opted for a single-day experience sampling approach because the application of the ear-EEG sensors required trained researchers, making it most practical for participants to visit the lab before returning home to complete the study.

\subsubsection{Natural Work Tasks}
In contrast to typical ESM studies that last several days or weeks (e.g., \cite{Rissler2020, Zueger2017, Zueger2018}), our study focused on collecting flow reports within a single day. To maximize data collection and focus on work-related flow, we limited the sampling to participants' work sessions instead of sampling during diverse activities throughout the day. This allowed us to balance task variety with frequent interruptions. During four one-hour sessions, participants worked on self-selected projects, and their flow experiences were probed at random intervals (after 10 or 20 minutes) three times per session, preventing anticipation of interruptions. 
To increase the chances of flow occurring, participants were asked to work on a project where they already had substantial expertise, but that still presented a challenge, replicating conditions known to support flow emergence. Furthermore, we restricted the scope of projects to two types: an academic thesis or a software engineering project — both of which have been shown to elicit high levels of flow in previous research \cite{Brown2023, Cowley2022, Knierim2018, Graft2023, Quinn2005}. These tasks were chosen to reflect common knowledge work activities that could provide a realistic, representative day of work, similar to designs used in prior studies \cite{yu2023semi, Graft2023, Rissler2020}. \\
 
To confirm the appropriateness of this task selection for eliciting flow, we also conducted a survey with 76 participants (all academic researchers and students, 24 female, age range 23-52 years, mean age 29 years) inquiring about their perceived likeliness of experiencing flow in a set of five work categories that was derived from previous field studies on flow in knowledge work \cite{Quinn2005}. Using the three-item short flow proneness at work scale \cite{Moneta2017}, survey respondents indicated the highest likeliness for experiencing flow during programming (mean = 3.46, SD = 0.99), followed by writing (mean = 3.37, SD = 0.79), data analysis (mean = 3.33, SD = 0.93), developing presentations (mean = 2.75, SD = 0.85), and task planning (mean = 2.74, SD = 0.95) with a value of 1 representing \textit{"never or almost never"} and 5 representing \textit{"always or almost always"}.

\subsubsection{Mental Arithmetic Task Reference}
The mental arithmetic task was chosen as a well-established flow induction approach \cite{Katahira2018, Ulrich2014, Knierim2018}, selected over other tasks like Tetris, Pacman, or chess due to their closer alignment with the cognitive demands of knowledge work. We specifically avoided using game tasks to prevent the potential influence of their hedonic aspects on the results.
In the mental arithmetic task, participants were asked to solve equations of varying difficulty as quickly as possible, with each trial lasting up to 18 seconds and followed by a four-second break. Three difficulty levels were used: "easy" (simple addition of 101 + [1, 2, or 3]), "optimal" (starting with two two-digit numbers, adjusting the number of summands based on performance), and "hard" (starting with four two-digit numbers, adding more after correct answers but not reducing them after mistakes). The adaptive difficulty in the "optimal" and "hard" conditions followed flow theory, ensuring optimal challenge in the former condition and creating a high, unmanageable difficulty in the latter condition \cite{Keller2016}.

\subsubsection{Procedure}
\begin{figure*}[h]
  \centering
  \includegraphics[width=.8\linewidth]{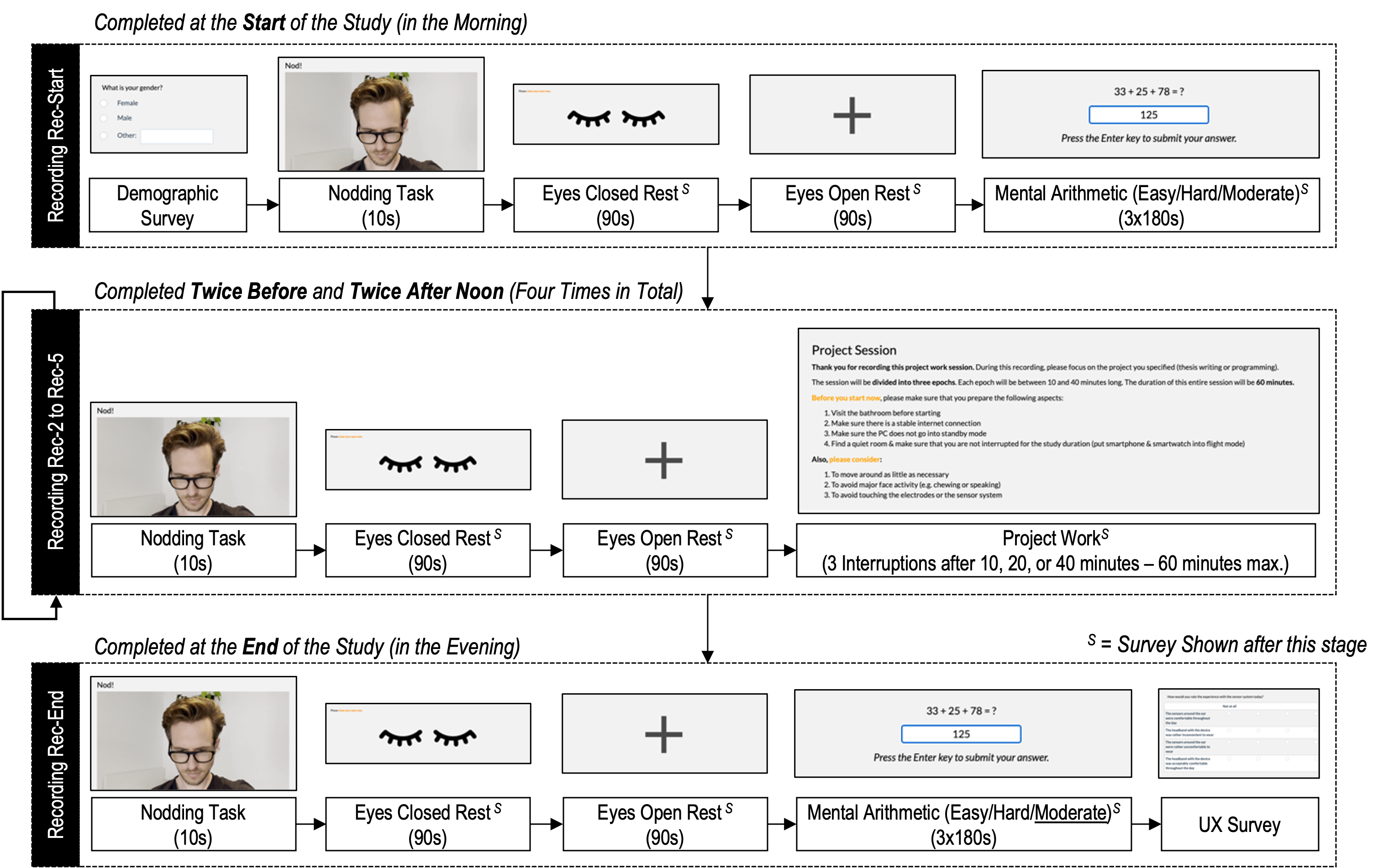}
  \caption{Visualization of the experiment procedure with screenshots from the various tasks and instructions. \revision{Top and bottom show the controlled math task processes, the middle shows the natural work task sessions that were repeated four times.}}
  \label{fig:studyProcedure}
  \Description{The image shows the experimental procedure for a study conducted over a full day, divided into three main phases: Rec-Start, Rec-2 to Rec-5, and Rec-End. The Rec-Start phase, completed in the morning, begins with participants filling out a Demographic Survey, followed by the Nodding Task (lasting 10 seconds), and two 90-second rest periods—Eyes Closed Rest and Eyes Open Rest. After these tasks, participants complete a Mental Arithmetic session (with Easy, Hard, or Moderate difficulty levels), performing three rounds of 180-second calculations, with a survey shown afterward. The second phase, Rec-2 to Rec-5, takes place Twice Before and Twice After Noon. In each session, participants again go through the Nodding Task and the two rest periods (Eyes Closed Rest and Eyes Open Rest, each for 90 seconds). In addition to these tasks, participants engage in a Project Session, which involves working on a project task (either thesis writing or programming) for a maximum of 60 minutes, with interruptions scheduled at the 10, 20, or 40-minute marks. Finally, the Rec-End phase occurs at the end of the day and mirrors the Rec-Start tasks. Participants complete the Nodding Task, the two rest periods, and a final Mental Arithmetic task. The study concludes with a UX Survey.}
\end{figure*}

The complete study procedure is visualized together with task screenshots in Figure \ref{fig:studyProcedure}. 
The participants were instructed to complete the mental arithmetic task sessions two times, once at the start of the study and once at the end of the study day.
The project work sessions were spaced out over the course of the day (i.e., participants received instructions to complete two work sessions before, and two sessions after noon, but could otherwise select the times freely). 
Before each recording, participants were instructed to ensure comparable and calm working conditions: visiting the bathroom, preparing required materials (devices, food, workspace, etc.), and minimizing distractions (closing the door and putting the smartphone in flight mode). 
Also, at the beginning of each recording (i.e., either mental arithmetic or project work), participants completed \revision{a nodding task for data synchronization} and eyes-open and eyes-closed resting phases to provide EEG baseline values. \revision{The comparison of eyes-open and eyes-closed data also provided a simple way for assessing signal recording reliability}.

\subsubsection{Survey Instruments}
After each task condition and work interruption, participants responded to surveys, capturing their perceived flow experience (Flow Short Scale FKS - 10 items \cite{Engeser2008}), mental workload (NASA TLX – 6 items \cite{Hart1988}), and affective state (one arousal and valence self-assessment manikin item \cite{Bradley1984}). Thereby, 18 report instances per participant were collected over the entire day (2 * 3 * mental arithmetic + 4 * 3 * project).
In addition, a survey at the beginning of the study was used to collect demographic data (gender, age, glasses, general health) and information on participants experience with their chosen project (measured as time spent on this and similar projects - 2 items from \cite{Erhard2014} and their self-reported flow proneness during this project (i.e., a metric of their tendency to experience flow in it - measured by the 3-item short flow proneness at work scale \cite{Moneta2017}). \\

\subsubsection{Wearable Ear-EEG}
Leveraging the properties of volume conduction \revision{(i.e., electrical signals spreading from their source accross the head to more remote sensor positions)}, we assumed that gelled ear-EEG could be a good option for studying flow in everyday life because previous work has already found flow correlates in EEG features that are visible across the scalp \cite{Ewing2016, Berta2013, Chanel2011} and in temporal regions in particular \cite{Wolf2014, Kramer2007, DeKock2014, Berta2013, Chanel2011}. 
Thus, to realize the wearable ear EEG recordings, we decided to use the ready-made cEEGrid electrode arrays that have been employed in numerous neuroscientific studies \cite{Debener2015, Wascher2019, Hoelle2021, Knierim2021BCI}. 
The cEEGrid ear-EEG electrodes have been shown to provide high signal quality and wearer comfort over the course of a day, while featuring a minimalistic appearance \cite{Debener2015, Hoelle2021}. The cEEGrids' recording quality is primarily realized by the possibility of using the electrodes with a gel enclosed by an adhesive sticker. Thereby, the gel remains fluid for long periods without causing discomfort or requiring the hair to be washed \cite{Debener2015}. The electrodes' application around the ear is realized in about five minutes. \\

\begin{figure*}[h]
  \centering
  \includegraphics[width=0.9\linewidth]{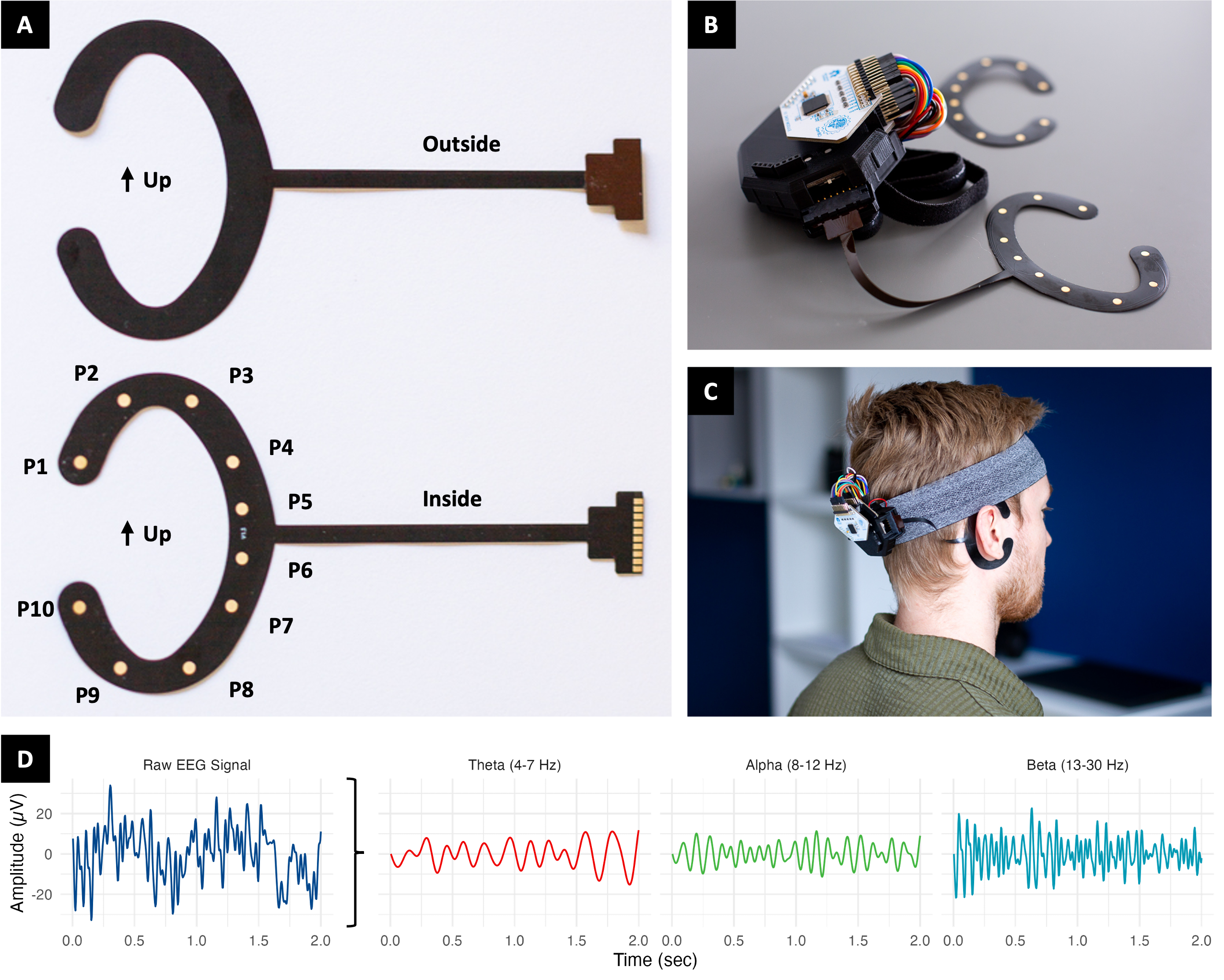}
  \caption{Gold-plated open-cEEGrids (A) together with OpenBCI Cyton+Daisy 16-channel biosignal amplifiers (B), worn by a study participant (C), \revision{and with examples of one channel's signals with decomposed frequency bands (D).}}
  \label{fig:ceegridsApplied}
  \Description{Left: A C-shaped open-cEEGrid electrode with 10 electrode positions labeled P1 through P10. This flexible electrode array is designed to fit behind the ear. Right: A side profile of a person wearing the sensing system on a headband, with the open-cEEGrid electrodes wrapped around the back of the ear. The headband holds a small, lightweight electronics module securely in place on the back of the head, ensuring continuous EEG recording without impeding the user's movement or comfort.}
\end{figure*}

In this study, we used the gold-plated cEEGrid version (open-cEEGrids), together with the low-cost, open-source Cyton+Daisy biosignal amplifiers from OpenBCI (NY, USA). We chose this system for its lower price point (which allowed us faster, simultaneous data collection with four participants a day), the low battery power consumption that ensured full-day recording coverage (see \cite{Rashid2018} for details - we used a 1000mAh LiPo battery, which allowed continuous recordings of over 12 hours), and the possibility to record data on an SD card which eliminated the need to implement software that the participants would have had to handle at home. With this kit, the amplifier is attached to a headband, as the cEEGrids electrodes are directly attached to the skin (see Figure \ref{fig:ceegridsApplied}). Here, we used the OpenBCI Cyton+Daisy board combination to collect electrophysiological recordings on 16 channels (plus reference and ground electrodes) at a sampling rate of 250Hz. Thereby, almost 200 hours of EEG data were collected.

\subsubsection{Implementation \& Sampling}
To ensure that participants could complete the study entirely self-directed and with minimal technical effort (i.e., without installations or recording system configurations) we devised a two staged study setup. First, participants were required to visit the laboratory in the morning between 8 and 9 AM to be fitted with the OpenBCI-cEEGrid system, and in the evening between 5 and 7 PM to return the device. 
The fitting session in the morning comprised an introduction and revision of the experiment and technology, the signing of an informed consent form and the application of the open-cEEGrids. The latter step included the preparation of the skin around the ear using alcohol swabs and an abrasive gel (Abralyt HiCl) to remove oily residues and old skin cells, which can lower signal-to-noise ratios (SNR). The open-cEEGrids were then applied using a lentil-sized drop of the same gel on each electrode. The electrodes' impedances were then assessed in OpenBCI GUI and confirmed to be below 40kOhm. Afterwards, the SD card recordings were started, and participants were sent home where they could conduct the rest of the study on their own accord.
Second, at home, participants could complete the experiment entirely through a browser-based interface that presented tasks, stimuli, scheduled task interruptions, and collected the self-reports. This experiment interface was implemented in LimeSurvey V.2.64.7 with the tasks being implemented as custom JavaScript programs. \\

\begin{figure}[h]
  \centering
  \includegraphics[width=1.0\linewidth]{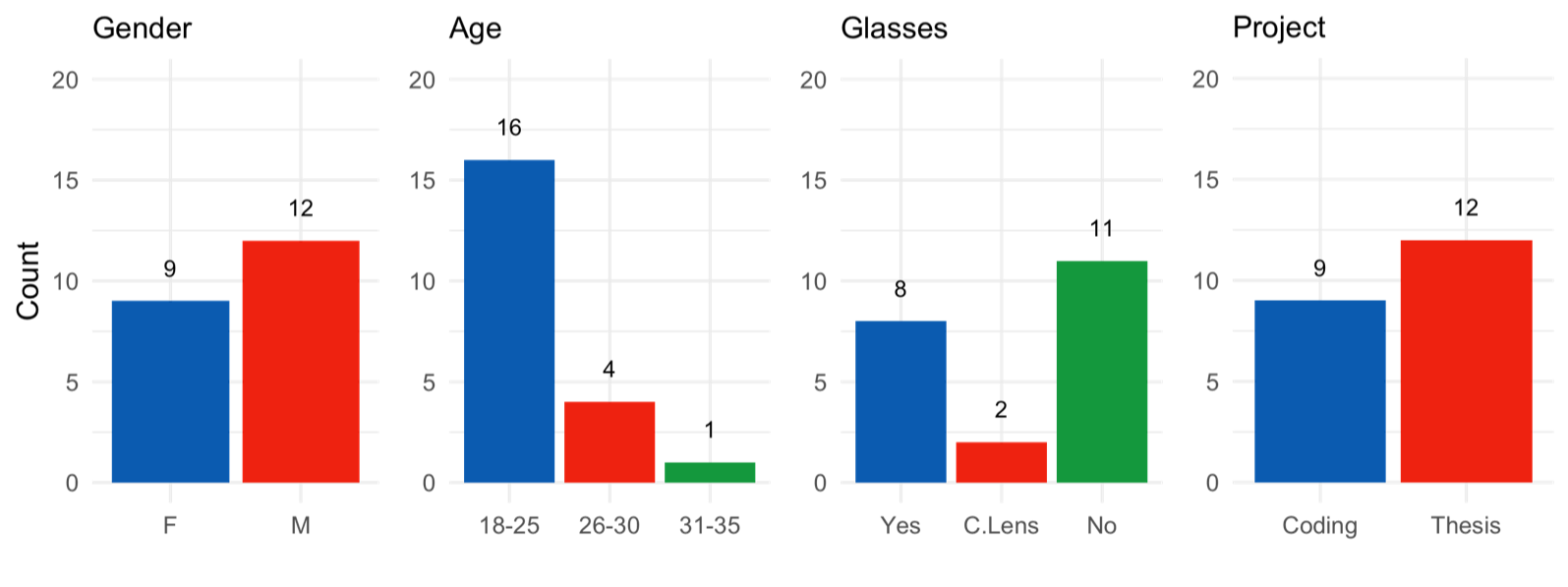}
  \caption{\revision{The study sample demographics show a relatively young group with balanced gender and project distributions.}}
  \label{fig:demographics}
  \Description{Demographic characteristics of the study sample. The first plot shows gender distribution, with two bars representing 9 females (F) and 12 males (M). The second plot shows the age distribution, with three bars indicating that 16 participants are between 18-25 years, 4 are between 26-30 years, and 1 participant is between 31-35 years. The third plot shows the participants' use of glasses or contact lenses. Eight participants wear glasses, 2 use contact lenses, and 11 do not wear any. The fourth plot represents the type of project participants are engaged in: 9 are coding, and 12 are working on a thesis project.}
\end{figure}

The study followed the Declaration of Helsinki and was registered with the authors' institutional review board. Participants received a compensation of 40 MU. This compensation amount was determined by a pre-test survey with 25 participants who were shown a description of the study (wearing the OpenBCI-cEEGrid system on a headband over the course of the day, visiting the lab before and after the study, and completing four sessions and additional controlled tasks). 
During the data collection, up to four participants were sampled a day – as four amplifiers were available. Participants were sampled at a large technical university and screened for being generally healthy \revision{(and not taking any mind-altering medication)}, willing to participate for the entire study duration, \revision{regularly working from home}, and being in the process of working on either an academic thesis or a software engineering project. \revision{Participants were screened to exclude those in the first or last 20\% of their project's timeline, based on their reported total and elapsed project weeks.}
Through the screening process, 22 individuals were recruited. One of these participants stopped participating in the study in the first two hours as it became apparent that working from home was not possible due to construction work in the neighboring apartment. The characteristics of the remaining 21 participants are shown in Figure \ref{fig:demographics}.

\subsection{Data Processing}
\subsubsection{Self-report Aggregation}
For the self-reports, multi-item self-report constructs’ internal consistency was assessed. 
For the flow construct (FKS), very good internal consistency was indicated (Cronbach's Alpha: 0.91, avg. item correlation: 0.49).
For the mental workload construct (NASA TLX), only after dropping the item on self-perceived task performance, the internal consistency was sufficiently improved for this construct as well (Cronbach’s Alpha: 0.86, avg. item correlation: 0.56).
Thus, the constructs were considered to show good coherence which is why they were subsequently aggregated (NASA TLX = sum of all 5 remaining items, FKS = mean of all 10 items).

\subsubsection{EEG Data Treatment}
To prepare the EEG data for analysis, first, data completeness was inspected (number of available recordings, channels, and samples), \revision{and signal reliability assessed.}
This process revealed that two EEG data sets had not been correctly written to the SD card and were thus not available. The cause was traced back to a faulty SD card. Furthermore, for two recordings of one amplifier, the connection to the Daisy board became unstable, during the day, which caused the loss of the right ear cEEGrid data. As the extracted features rely on data from both sides of the head, these two datasets were also unusable. 
In total, 17 complete EEG recordings were available for further study. 
\revision{Previous work has shown the suitability of gelled cEEGrids for full-day use \cite{Debener2015}. Nevertheless, to confirm reliable neural signal collection, we tested for the Berger effect (increased alpha power with eyes closed) at the start of each session. Higher alpha power was consistently observed in the eyes-closed condition, and a linear mixed model (random intercepts for participants) found no effect of session repetition (accumulated sessions) on relative alpha power differences (EC-EO, p=0.2611), indicating stable EEG data quality throughout the day.} \\

Next, the EEG data of interest were cut, shortening the extraction window by five seconds on each side, ensuring that the used EEG data corresponded to the selected study task. Thus, EEG data from 170-second windows of the mental arithmetic task conditions, and 170-second windows before each project work interruption were extracted (i.e., 18 instances per participant). 
Each of these segments were then subjected to line noise removal using the ZapLine algorithm and bandpass-filtered (2-30 Hz FIR - \revision{default parameters in mne python's filter\_data function() - V1.8.0}) to remove slow signal drifts and high-frequency noise. Afterwards, the signal data were re-referenced to the average of L5 and R5 electrodes (behind the ear – similar to previous cEEGrid research \cite{Debener2015, Wascher2019, Knierim2021BCI}, mimicking a robust re-referencing approach \cite{Bigdely2015}). This means that bad channels were first identified and interpolated to ensure that the re-referencing step did not introduce additional noise to the remaining channels. Bad channels were identified using the artefact subspace reconstruction (ASR) routine, which determines bad channels based on low correlations with neighboring electrodes \cite{Mullen2015}. Next, the ASR routine was used again to clean the data further and, particularly, repair non-stationary artefact bursts by interpolating noisy epochs based on clean data segments \revision{(from the eyes-open rest stages at the start of each session)}. Riemannian geometry was used, as the rASR method has shown better performance for cleaning artefacts from mobile EEG data \cite{blum2019}. 
For each electrode and study segment, we then extracted 1Hz frequency band powers by computing the power spectral density using a Welch-windowed Fast Fourier Transformation (FFT) using 2-second window widths without overlap. The resulting \revision{powers in each frequency bin} were normalized by dividing them by the total power from all frequencies. This step was included to provide a relative measure of frequency band activity that is more comparable within and between participants. Next, the electrodes were \revision{mean} aggregated to regions of interest (ROI): left and right ear and all channels combined.
Finally, the remaining frequencies were then mean aggregated into standard frequency bands (theta: 4–7 Hz, alpha: 8–12 Hz, and beta: 13–30 Hz) and the hemispheric asymmetries were calculated by subtracting the left from the right ear band powers in the frequency bands.

\section{Results}
\subsection{Flow in Single-Day Knowledge Work}
\begin{figure*}[h]
  \centering
  \includegraphics[width=.9\linewidth]{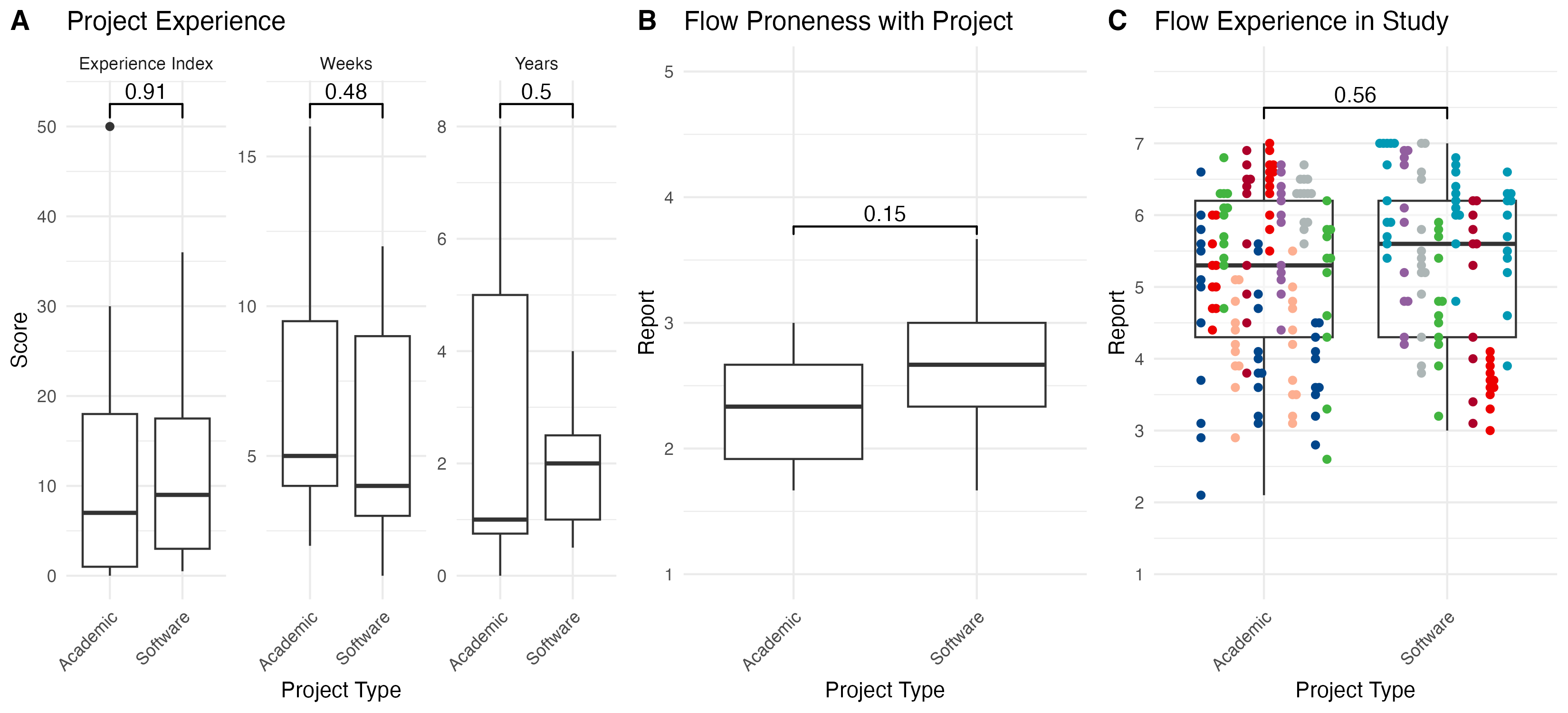}
  \caption{Distributions and test results for project type experience (A), flow proneness with the project (B), and actual flow experience during the study by project type and participant (C). Numbers above bars represent the p-values of the corresponding statistical tests. \revision{No significant differences are found in all variables.}}
  \label{fig:FlowAtWork}
  \Description{A 3-part figure. Figure A: Project experience across academic and programming task types. Boxplots display scores for three different metrics: experience index, weeks, and years. The x-axis displays the project type (academic or programming), and the y-axis measures the score for each metric. The experience index shows a similar distribution for both task types, while the programming project type shows a slightly higher score for years of experience, and the academic projects a slightly higher score for weeks. None of the differences are statistically significant though. Figure B: flow proneness reported by participants in academic and programming tasks. The y-axis represents the reported flow proneness, with values ranging from 1 to 5. The programming task type shows a slightly higher median and overall distribution compared to academic tasks, although the difference is not statistically significant as shown by the p-value (0.15) labeled above the boxplot. Figure C: flow experience during the study for academic and programming tasks. Boxplots show the reported flow experiences, with individual data points color-coded. Both task types exhibit a range of flow experiences, with programming tasks showing a slightly higher median. A non-significant difference is indicated by the p-value (0.56) between the task types.}
\end{figure*}

Before investigating the emergence of flow in the natural work tasks, we assessed whether both project types provided similar opportunities for it.
Adapting a method commonly used in music and creative writing research \cite{Erhard2014}, we derived an individuals' project type experience index by multiplying general project experience (measured in years) with experience on the current project (measured in weeks spent on the project). The individual reports and the index are shown in Figure \ref{fig:FlowAtWork}A. Independent Welch t-Tests show no significant difference in experience levels between the two project types (experience index: t=0.1208, p=0.9052, years of experience: t=0.6950, p=0.4973, weeks in project: t=0.7217, p=0.4799).
Also, participants' reports showed no significant difference for flow proneness in their project (t=-1.5029, p=0.1543, see Figure \ref{fig:FlowAtWork}B). Therefore, both project types appear to provide similar pre-conditions for flow emergence. \\

Next, to answer RQ1 (i.e., to learn how flow experiences emerge during various forms of natural knowledge work), we assessed the distribution of flow reports across the two project types (see Figure \ref{fig:FlowAtWork}C). With median flow reports above five on the 7P Likert scale, the intensity of the reported flow can be considered relatively high, when comparing it with related work \cite{Quinn2005, Katahira2018, Knierim2018} and when comparing it to the controlled flow induction task (see below).
Further, we found no significant difference in the average flow intensities across the two work project types (ANOVA on a LMM with participant as random intercept: F=0.3605, p=0.5557, EMM Academic Thesis=5.11, Software Engineering Project=5.35), and no significant difference in the range (max-min) of flow (ANOVA on LMM with participant as random intercept: F=2.5038, p=0.1286, EMM Academic Thesis=1.17, Software Engineering Project=0.81).
Thus, there was no clear difference between the two project types in eliciting flow. \\

\begin{figure*}[h]
  \centering
  \includegraphics[width=1.0\linewidth]{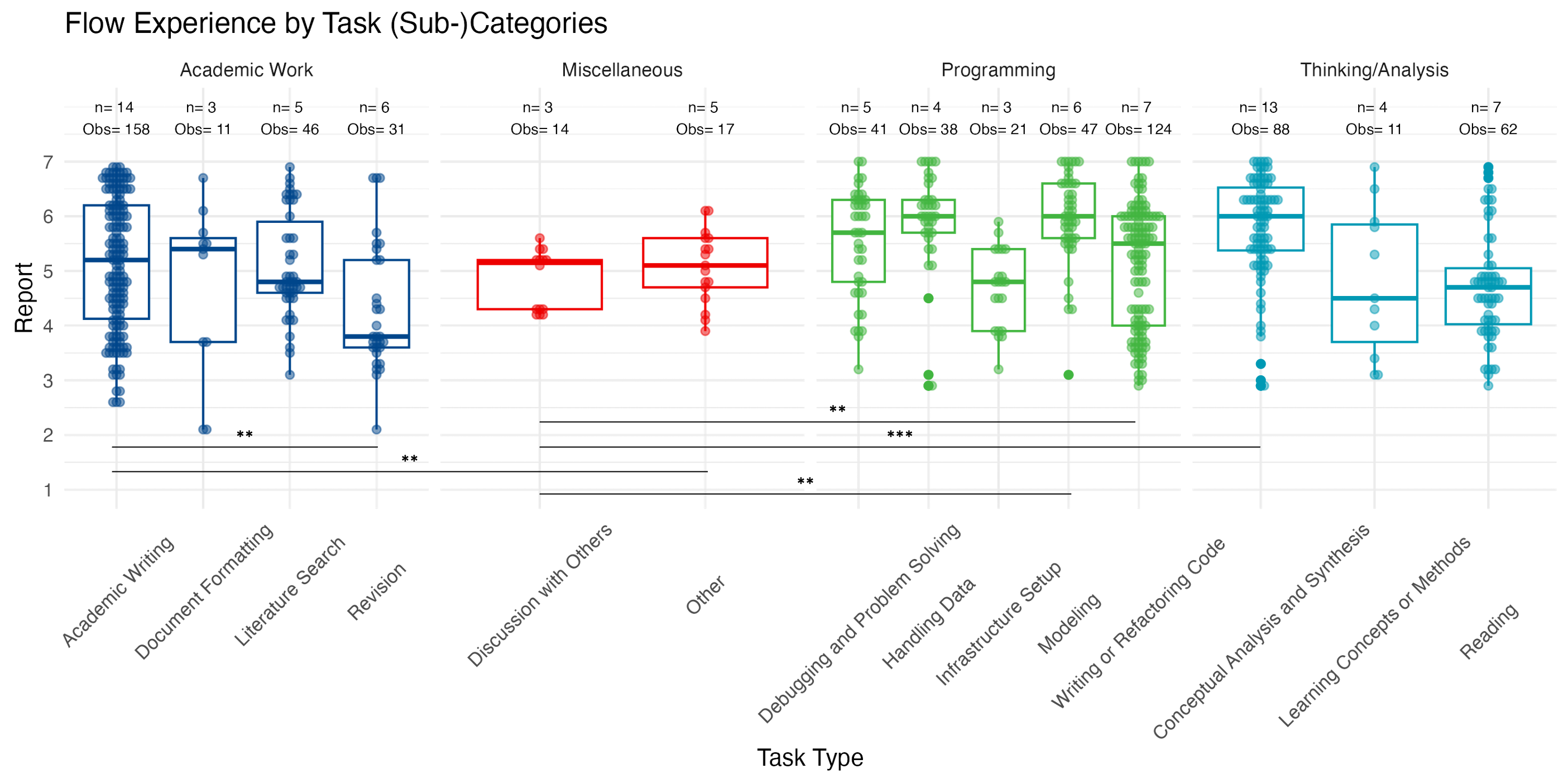}
  \caption{Distribution of flow in the reported activities during the natural work sessions. \revision{The majority classes are academic work, programming, and thining/analysis.} Numbers above the box plots show the number of observations in total per sub-category and for how many participants the category occurred. \revision{Estimated marginal mean differences with p<0.01 are shown with bars.}}
  \label{fig:taskDescriptions}
  \Description{The figure presents the distribution of flow experiences across different task subcategories, organized into four main categories: Academic Work, Miscellaneous, Programming, and Thinking/Analysis. The x-axis lists the subcategories under each main category, while the y-axis shows flow reports on a scale from 1 to 7. Each boxplot represents the range of flow experiences for a given subcategory. In the Academic Work category, tasks such as Academic Writing, Document Formatting, Literature Search, and Revision exhibit varied flow experiences, with Academic Writing showing by far the most observations. The Miscellaneous category includes subcategories like Discussion with Others and Other. This category in general doesn't feature many observations, compared to the other three. The Programming category contains subcategories like Debugging and Problem Solving, Handling Data, Infrastructure Setup, Modeling, and Writing or Refactoring Code. Lastly, in the Thinking/Analysis category, tasks like Conceptual Analysis and Synthesis, Learning Concepts or Methods, and Reading show a range of flow experiences. Conceptual Analysis and Synthesis seems to have evoked stronger flow experiences than the other two sub-categories.}
\end{figure*}

To further examine the flow experiences, it is important to understand, what participants were actually doing in their self-selected natural work sessions. To that end, participants answered the prompt \textit{"Describe in one sentence what you were doing just before the survey came up"} at each work session interruption. 
A thematic analysis was conducted on these responses, using an inductive coding approach~\cite{Braun2006}. 
Two researchers coded half of the participant responses independently, sampling participants randomly. Duplicates were expelled, and a final coding tree was jointly developed and refined through an in-depth discussion of results. Subsequently, one researcher coded the rest of the responses. Based on the coding tree, the responses were clustered into four main work categories (Academic Work, Programming, Thinking/Analysis, and Miscellaneous) with multiple sub-categories (14 in total). For all four categories, a large subcategory class emerged that contained rather abstract descriptions (responses like "Coding" - P19, or "Writing a text" - P9). More detailed responses, were clustered into the remaining sub-categories.
The distribution of the reported occurrences in Figure \ref{fig:taskDescriptions} and exemplary statements for each category in Table \ref{table:taskDescriptions} show that the large majority of the activity can be considered programming (34.83\%), academic work (33.71\%), or falling into a more abstract category related to both tasks: thinking/analysis (26.59\%). Only 4.87\% is considered miscellaneous activity, predominantly discussion with others. \\

\begin{table*}[h]
\caption{Examples for work session reports and associated categories.}
\centering
\begin{tabular}{l p{0.6\textwidth}}
\textbf{Category} & \textbf{Quote} \vspace{2pt} \\ \hline
\textbf{\textit{Programming}} &\\
\hline
Debugging and Problem Solving &\textit{ "Looking for a bug in my code concerning database connection" – P1 }\\
Handling Data & \textit{"Parsing the Waymo dataset with Python" – P20 }\\
Infrastructure Setup &\textit{ "Rebuilding a library" – P22} \\
Modeling & \textit{"Combining previously implemented and trained models into one" – P18 }\\
Writing or Refactoring Code &\textit{ "Writing more code" – P10 }\vspace{2pt}\\
\textbf{\textit{Academic Work}} & \\
\hline
Academic Writing & \textit{ "I summarized the results of an experimental study" – P7} \\
Document Formatting & \textit{"Setting up a LaTeX document" – P15 }\\
Literature Search & \textit{"Searching in ScienceDirect" – P12 }\\
Revision & \textit{"I did some corrections in my written text" – P4 }\vspace{2pt}\\
\textbf{\textit{Thinking/Analysis}} & \\
\hline
Conceptual Analysis and Synthesis & \textit{"Comparing results of different papers and writing them down" – P7 }\\
Learning Concepts or Methods & \textit{ "Watching tutorial" – P21 }\\
Reading & \textit{ "Reading online code documentation" – P22 }\vspace{2pt}\\
\textbf{\textit{Miscellaneous}} & \\
\hline
Discussion with Others & \textit{"Got a call from a coworker working on the same project" – P21 }\\
Other & \textit{ "Looking for words in the translator" – P12 }\\ \hline
\end{tabular}
\label{table:taskDescriptions}
\end{table*}

\begin{figure*}[h]
  \centering
  \includegraphics[width=0.9\linewidth]{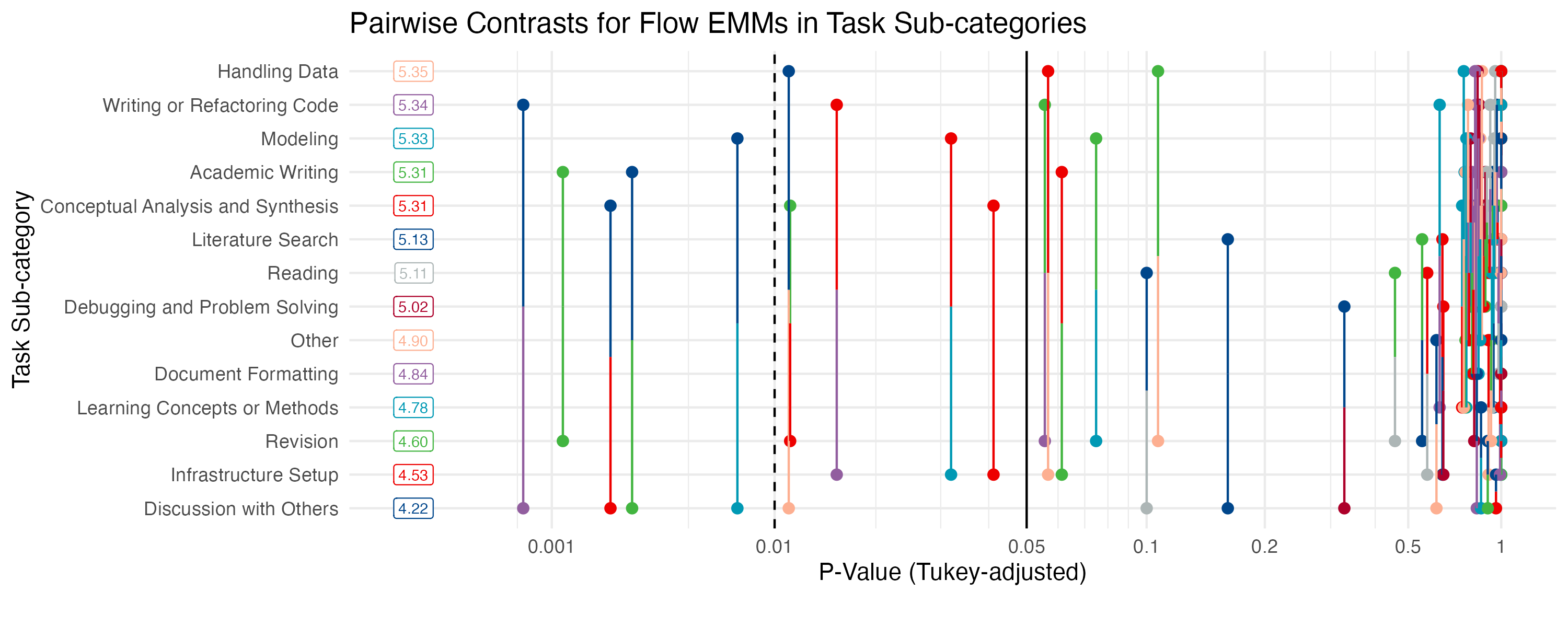}
  \caption{Pairwise post-hoc contrasts for flow experience by work category sub-tasks based on an LMM. \revision{The x-axis shows Tukey-adjusted p-value levels with the common .01 and .05 cut-off levels as vertical lines.} Significant differences emerge between the more intense flow situations (Writing code, Handling data, Modeling, Academic Writing, Conceptual Analysis and Synthesis) with the lower flow situations (Revision, Infrastructure Setup, Discussion with Others).}
  \label{fig:flowByCategoryPWPP}
  \Description{The plot illustrates pairwise contrasts for Estimated Marginal Means (EMMs) of flow reports across different task sub-categories. The y-axis lists task sub-categories such as Handling Data, Academic Writing, and Debugging, among others. Each sub-category has its EMM shown on the left, and these values range from 3.11 for "Reading" to 5.54 for "Handling Data." The x-axis represents p-values ranging from 0.001 to 1.}
\end{figure*}

An ANOVA based on a linear mixed model (LMM) with the work sub-categories as the fixed effect and participants as a random intercept showed a significant difference for flow (F=5.0754, p<0.0001), indicating that some sub-category tasks elicited stronger flow than others.
%
Follow-up pairwise contrasts based on estimated marginal means (EMMs) are shown in Figure \ref{fig:flowByCategoryPWPP}. Here it can be seen, that significant differences emerge between the more intense flow situations (Writing code, Handling data, Modeling, Academic Writing, Conceptual Analysis and Synthesis) with the lower flow situations (Revision, Infrastructure Setup, Discussion with Others). While there are no significant differences between the major academic thesis or software engineering sub-tasks, the coding tasks show a slight tendency towards more intense flow. \\

\revision{Altogether, we found that both academic thesis writing and software engineering projects offered similar opportunities for flow (based on project experience and flow proneness) with comparable flow intensities and ranges. Participants performed diverse knowledge work tasks consistent with prior field studies on flow in knowledge work~\cite{Quinn2005, Kurosaka2023}, providing a solid basis for comparing flow across paradigms using physiological sensors.}

\subsection{Flow Across Paradigms}
To answer RQ2 (i.e., to compare how natural flow experience compares to flow experiences in a classic lab paradigm), we inspected the flow reports by task paradigm. 
We focused on two main analyses that follow-up questions from the literature: a) do natural tasks elicit more intense flow, and b) do controlled tasks create better experience contrasts? \\

Thus, to answer RQ2a, we first filtered the data for only the most intense flow experiences in each recording session. 
\revision{For each participant, the two most flow-inducing situations from the math task and the four most flow-inducing situations from the natural work tasks were selected (based on the self-reports values). This filtering ensured a fair comparison, as the math task includes low-flow situations by design, which could otherwise bias results in favor of the natural work task.}
We set up an LMM with flow as dependent variable, the task paradigm (math task vs. natural work) as dependent variable, and a random intercept for each participant. An ANOVA test of this model showed that the natural work elicits significantly more intense flow experiences (F=4.9225 p=0.0284; EMM Math = 5.51, Natural Work = 5.74). \\

Next, to answer RQ2b, comparing the range (max-min) of flow intensities per paradigm and session (ANOVA on LMM with participant as random intercepts: F=36.9337, p<0.0001, EMM Math=1.88, Work=1.03) shows that flow intensities vary more strongly in the artificial math tasks than the natural work tasks. To further evaluate, we also tested whether the total range in a participant's experienced flow is greater without or with the math task data (paired t-test: t=4.6148, p=0.0001) and found that the experienced flow experience range across participants is significantly lower without the math task data on average (mean range work alone = 2.3174, mean range work+math = 3.3174), and lower for all but one participant. This indicates that in natural work, we don't find the same range of flow experiences as known from controlled task paradigms. Importantly, the natural work tasks appeared to lack low flow experience situations. \\

\begin{figure*}[h]
  \centering
  \includegraphics[width=1.0\linewidth]{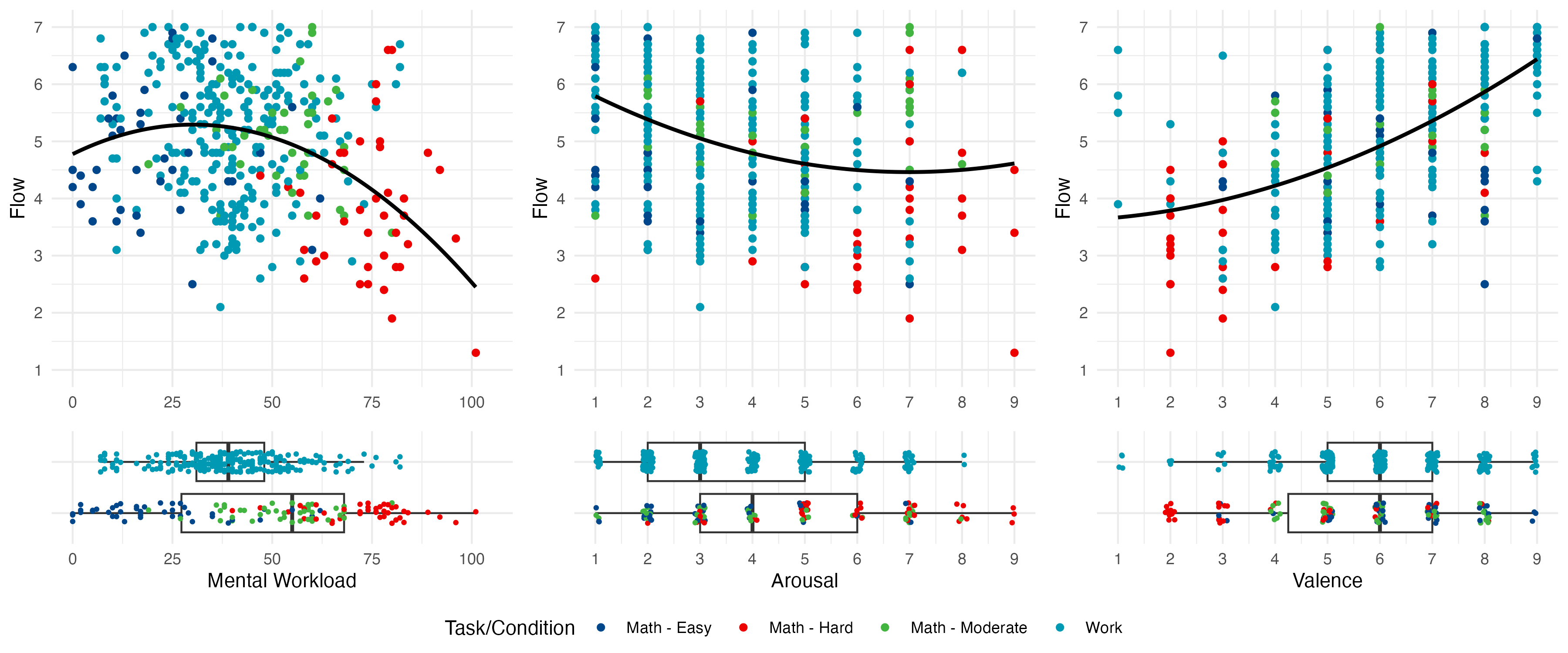}
  \caption{\revision{Flow shows significant quadratic relationships with mental workload, arousal, and valence. Lines are fitted using quadratic mixed models with participants as random intercept. The boxplots on the bottom show reports per task (math or natural work).}}
  \label{fig:FlowAndCo}
  \Description{This figure consists of six plots arranged in a 2x3 grid. Each plot visualizes relationships or distributions of data points associated with a variable labeled "Task." The plots use a combination of scatter points and boxplots, with points color-coded into three categories (blue, green, and red).
  In the top-left plot, a scatter plot displays data points spread across a horizontal axis ranging from 0 to 100 and a vertical axis ranging from 1 to 7. The points form clusters, and a black curve is overlaid to indicate a fitted trend line. The top-center plot uses a similar scatter plot format but with fewer horizontal categories (ranging from 1 to 9). The points cluster vertically above each horizontal value, and the trend line spans these categories. The top-right plot simplifies this further, showing only two horizontal categories with clustered points and a trend line connecting them.
  The bottom-left plot visualizes the data as horizontal strips where points cluster along the horizontal axis from 0 to 100. The vertical axis is binary (0 or 1), and points are grouped above these levels. The bottom-center plot combines a boxplot with overlaid scatter points for horizontal categories ranging from 1 to 9. The box-and-whisker structures summarize the central tendencies and spread, while the points show individual values. The bottom-right plot uses a similar format, but the horizontal axis includes only two categories. The boxplots summarize the data distributions, with points overlaid to show individual data values.
  Overall, the figure captures how the distribution of the "Task" variable varies with different horizontal variables and categories.}
\end{figure*}

\revision{We thus investigated why there might be fewer "low flow" situations in work, assessing the distributions of workload, arousal, and valence reports, variables that have a known relationship with flow~\cite{Knierim2021BCI, Nacke2008, Chanel2011, Ewing2016, Fairclough2013}.
Given the reported non-linear relationship between flow and workload~\cite{Knierim2021BCI}, quadratic mixed models with participant random intercepts were used. Based on the predicted values (see Figure \ref{fig:FlowAndCo}), we see an inverted-U shaped relationship between flow and workload (linear term: b=-6.115, p<0.001, quadratic term: b=-5.604, p<0.001), and U-shaped relationships with arousal (linear term: b=-7.757, p<0.001, quadratic term: b=2.907, p=0.005), and valence (linear term: b=13.140, p<0.001, quadratic term: b=2.584, p=0.012). 
Lower flow was most common during high workload, high arousal, and low (negative) valence, particularly in the hard math tasks. The math task also showed a broader range of demand and affective responses, further highlighting smaller experience contrasts between the two task types.} \\

\revision{In summary, these results highlight substantial flow elicitation differences across the two paradigms. Natural work tasks elicited more intense flow but with a narrower range and no low-flow situations. In contrast, the controlled math tasks produced broader experience contrasts, with lower flow linked to high workload, arousal, and low valence, highlighting the controlled paradigm's ability to elicit a wider spectrum of flow experiences.}

\subsection{Ear-EEG Patterns} 
To answer RQ3 (i.e., to investigate how around-the-ear EEG can inform flow monitoring in natural knowledge work), we looked for known relationships of EEG frequency band power features and self-reported flow experience. 
The features were used as predictors in quadratic ordinary least squares (OLS) regression models \revision{with two-tailed tests} (see Table~\ref{tab:flow_ols_results}) to detect linear and non-linear relationships that have been related to flow in previous flow EEG research \cite{Ewing2016, Berta2013, Katahira2018, Fairclough2013, Chanel2011, Nacke2008}. To avoid issues with multicollinearity (adjacent EEG frequency bands are typically highly correlated), we set up separate models for the band power features. \\  

\begin{figure*}[h]
    \centering
    \includegraphics[width=\linewidth]{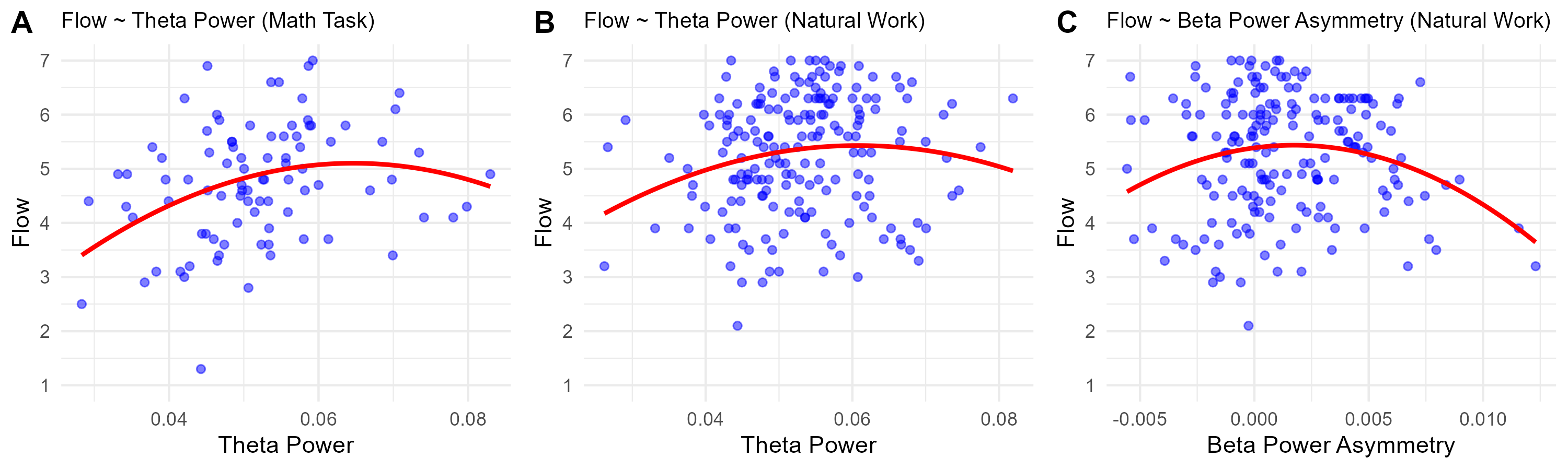}
    \caption{\revision{Flow experience shows significant quadratic relationships with theta and beta asymmetry frequency band powers. Lines are fitted using quadratic regression models.}}
    \Description{Three plots side by side. A) Scatter plot of flow vs. theta Power in a mental arithmetic task. Theta Power is plotted on the x-axis from 0.03 to 0.08, and flow is plotted on the y-axis from 1 to 7. Data points show an initial rise in flow as Theta Power increases, peaking around 0.05 and then falling, forming an inverse-u-shaped curve. A fitted red line represents this quadratic relationship. B) Scatter plot of flow vs. Theta Power in natural work tasks. Theta Power is plotted on the x-axis from 0.03 to 0.08, and flow is plotted on the y-axis from 1 to 7. Data points show a slight initial rise in flow with increasing Theta Power, followed by a decline, forming a negative quadratic trend. A red fitted line represents the negative quadratic relationship. C) Scatter plot of flow vs. Beta Power Asymmetry in natural work tasks. Beta Power Asymmetry is plotted on the x-axis from roughly -0.005 to 0.01, and flow is plotted on the y-axis from 1 to 7. Data points indicate that as Beta Power Asymmetry increases, flow initially rises, peaking slightly below 0.002, and then gradually declines, forming a negative quadratic-shaped trend. A fitted red line shows the quadratic relationship between flow and Beta Power Asymmetry. The points are widely distributed, with a concentration around flow values of 5 and Beta Power Asymmetry values near 0.}
    \label{fig:Scatterplots}
\end{figure*}

Similar to previous research, we find a statistically significant, \revision{quadratic} relationship predicting flow experience from theta power in the mental arithmetic task (see Figure \ref{fig:Scatterplots}A) and the natural work task (see Figure \ref{fig:Scatterplots}B). It should be noted, though, that in natural work, the quadratic term was significant at trend level only, which could be due to the lower presence of \revision{very easy and} very demanding situations \revision{(see Figure \ref{fig:FlowAndCo})} that might similarly cause disengagement.
\revision{The negative quadratic coefficients, together with the fitted curves indicate an inverted-U shaped relationship. Two-line tests \cite{Simonsohn2018} show that theta power exhibited a significant positive relationship with flow below the inflection point (math: inflection point x$_{c}$=0.060, slope=56.15, p=0.0007, natural work: inflection point x$_{c}$=0.060, slope=44.57, p=0.0014), with large slope changes but non-significant negative relationships after the inflection point (math: slope=-23.21, p=0.572, slope change=-79.36, natural work: slope=-11.57, p=0.718, slope change=-56.14).} \\

\begin{table*}[h]
\centering
\caption{Model results for \revision{individual} OLS regressions of flow and frequency band power features.}
\label{tab:flow_ols_results}
\begin{tabular}{l|rrrr|rrrr}
\toprule
& \multicolumn{4}{c|}{\textbf{Math Task}} & \multicolumn{4}{c}{\textbf{Natural Work}} \\
\cmidrule(lr){2-5} \cmidrule(lr){6-9}
EEG Feature & Estimate & Std. Error & t-value & p-value & Estimate & Std. Error & t-value & p-value \\
\midrule
Alpha & 61.886 & 63.216 & 0.979 & 0.330 & 73.175 & 52.625 & 1.390 & 0.166 \\
Alpha$^2$ & -1018.999 & 851.907 & -1.196 & 0.235 & -853.947 & 810.601 & -1.053 & 0.294 \\
\midrule
Beta & -253.421 & 141.674 & -1.789 & 0.077 & 14.787 & 85.259 & 0.173 & 0.863 \\
Beta$^2$ & 5357.214 & 3779.036 & 1.418 & 0.160 & -663.294 & 2460.504 & -0.270 & 0.788 \\
\midrule
Theta & \textbf{166.259} & \textbf{68.965} & \textbf{2.411} & \textbf{0.018} &  
\textbf{127.677} & \textbf{63.649} & \textbf{2.006} & \textbf{0.046} \\
Theta$^2$ & \textbf{-1285.284} & \textbf{624.338} & \textbf{-2.059} & \textbf{0.043} & -1050.252 & 585.728 & -1.793 & 0.075 \\
\midrule
Alpha Asym. & 14.256 & 24.587 & 0.580 & 0.564 & 14.915 & 18.642 & 0.800 & 0.425 \\
Alpha Asym.$^2$ & -2360.006 & 3857.522 & -0.612 & 0.542 & 2700.209 & 2285.481 & 1.181 & 0.239 \\
\midrule
Beta Asym. & 44.680 & 36.360 & 1.229 & 0.222 & 56.068 & 33.798 & 1.659 & 0.099 \\
Beta Asym.$^2$ & -3737.636 & 7553.278 & -0.495 & 0.622 & \textbf{-16085.257} & \textbf{5298.838} & \textbf{-3.036} & \textbf{0.003} \\
\midrule
Theta Asym. & -8.851 & 18.757 & -0.472 & 0.638 & -6.411 & 16.527 & -0.388 & 0.699 \\
Theta Asym.$^2$ & -1090.239 & 1940.553 & -0.562 & 0.576 & 98.717 & 1426.707 & 0.069 & 0.945 \\
\bottomrule
\end{tabular}
\end{table*}

While previous observations of theta \cite{Kramer2007, DeKock2014, Wolf2014} and alpha asymmetry \cite{Labonte2016} with flow are not confirmed, we find a novel convex quadratic relationship of beta power asymmetry and flow in natural work, with the peak of the quadratic model occurring around an asymmetry value of zero. This indicates that flow is highest when neither left- nor right-hemispheric beta activity dominates (see Figure \ref{fig:Scatterplots}C). \revision{Further, two-lines tests show a non-significant, positive relationship with flow below the inflection point (x$_{c}$=.001, slope=105.46, p=0.150), with
with a large slope change and a significant negative relationship after the inflection point (slope=-124.29, p=0.010, slope change=-229.75).}
To further understand the relationship, we conducted a set of follow-up analyses.
First, to assess whether the relationship is fully related to flow, we tested the relationship with the two sub-dimensions of the used flow scale (FKS): fluency and absorption.
Beta power asymmetry shows a significant quadratic relationship with both fluency (quadratic: b=-13,560, p=0.0097) and absorption (quadratic: b=-13,250, p=0.0223).
Next, we tested whether the beta asymmetry could be related to other, confounding variables that have been associated with flow. No significant relationships are found between beta asymmetry and valence (linear: b=-54.264, p=0.205, quadratic: b=8364.394, p=0.211), arousal (linear: b=75.028, p=0.145, quadratic: b=9766.026, p=0.224), or mental workload (linear: b=24.3728, p=0.961, quadratic: b=-15790, p=0.840).
Finally, to assess whether beta asymmetry might be related to project types, we compared the beta power asymmetry between academic thesis and software engineering projects and found no significant differences (LMM with participants as random intercepts, F=0.012, p=0.913, EMM Academic Thesis = 0.114, EMM Software Engineering Project = 0.207). \\

\revision{Altogether, we find that the well-known quadratic theta-flow lab finding  can be transferred to a natural field setting, bridging a critical gap in flow research, albeit the effect was stronger in the controlled task. Furthermore, we find a relationship with beta power asymmetry that was robust across flow sub-dimensions but not significantly associated with valence, arousal, mental workload, or task type, suggesting a novel, task-independent indicator of natural flow.}

\section{Discussion, Limitations, \& Future Work}
\subsection{Flow in a Single-Day Home-Office Setting}
Our study successfully elicited strong flow experiences by designing task conditions that were both challenging and aligned with participants' expertise in two common knowledge work settings. \revision{Also, as participants brought their own projects, genuine task motivation and external validity were ensured.} This approach offered an effective, efficient method for capturing flow within a single day of natural work, providing a foundation for future physiological field studies.
Both the academic thesis writing and the software engineering projects effectively induced flow. This is noteworthy, as prior research has often focused on software engineers, assuming that structured tasks with clear, frequent feedback cycles are particularly conducive to flow in knowledge work \cite{Brown2023, Cowley2022, Rissler2020}. However, alternatives are needed, as software engineering requires specialized skills not broadly developed in the general population. 
\revision{We added academic thesis writing as a representative knowledge work case, given the likely expertise and availability of our participant pool with it. However, future work could also explore even more common task alternatives like reading or email writing~\cite{Quinn2005, Kurosaka2023}, which might accommodate an even broader participant base.}
Our study descriptively (though not statistically) suggests that coding tasks may be more conducive to flow, with academic writing having a similar effect among students. While these results may not generalize to professional software engineers or researchers, they show that student populations can be valuable for advancing flow research in natural work settings.\\

At the same time, we observed relatively weak flow contrasts in the natural work tasks (i.e., a lower variance in flow reports). This could challenge physiological research, which often relies on more pronounced task differences (e.g., clearly easy or hard tasks) \cite{Ewing2016}. This limitation may have influenced our ear-EEG results, as the quadratic theta band relationship was weaker in natural work than in the mental arithmetic task. A possible explanation is that less variation in task demand or engagement limited this effect, with particularly low flow instances being rarer in natural tasks and more common in overly difficult controlled tasks \cite{Ewing2016, Keller2016}. \revision{This proposition is supported by the observerd narrower workload and valence experience ranges in the natural work.}
Other sensing approaches may face similar challenges, so future research could address this by designing single-day task programs that remain natural but vary in difficulty or engagement. For example, tasks could include different stages of a knowledge work project like starting, advancing, and completing it. Our sub-category analysis supports this, as revision tasks were less flow-conducive and could be used for such manipulations. Also, previous research has, for example, indicated that starting a new project can be very difficult and overwhelming \cite{Knierim2018}. \revision{To that end, it is possible, that focusing on mid-project work may explain the low to moderate workload levels, as initial and final stages are often more demanding in knowledge work projects.}
Alternatively, participants might be instructed to select tasks that they find more or less challenging within their ongoing project. Adding other manipulations to the study design could also help, like creating situations that might limit flow occurrence, for example, by adding environmental noise that creates interruptions \cite{Zueger2017, Zueger2018} or by enforcing multitasking \cite{Peifer2019}, as both influences have been found to substantially reduce flow.

\subsection{Flow Experiences Across the Two Paradigms}
This is the first study to directly compare flow research paradigms — natural task observation and controlled flow induction — using the same participants and repeated measures within a single day. We included this comparison to provide a reference for natural work tasks, selecting a mental arithmetic task commonly used in flow physiology research \cite{Katahira2018, Ulrich2014, Knierim2021BCI} that focuses on cognitive processes without a hedonic component. Our findings reveal significant differences in flow intensities and contrasts between the two approaches.
We found that natural tasks elicited more intense flow experiences than controlled lab tasks, supporting critiques that artificial tasks may limit flow intensity due to reduced intrinsic motivation and interference with attentional processes that are needed for experiencing the holistic task focus required for flow \cite{Hommel2010, Harris2017}. Flow theory also suggests that flow intensifies when a well-developed skill is applied to a challenging task \cite{Moneta2012}. Thus, the observed differences might also be attributed to the higher level of skill developed in the natural work tasks. Future research should compare controlled and natural tasks with comparable skill levels to better understand these dynamics. \\

Second, we should emphasize that we used an artificial task also employed in stress research \cite{Dimitriev2020}, which has been critiqued for its ability to elicit flow compared to game-based tasks using similar difficulty manipulations \cite{Keller2016}. We chose the mental arithmetic task for its simplicity and its conceptual and neural similarity to natural work projects, but it’s possible that gamified tasks could better complement natural flow research, as games are also a common source of natural flow experiences \cite{Klarkowski2015, Klarkowski2016}. Future studies could combine difficulty-manipulated games with natural gaming to further explore flow physiology in everyday life.


\subsection{Ear-EEG Patterns}
As the first study to employ ear-EEG measurements for monitoring flow in everyday settings, our work successfully replicates well-known flow correlates. Specifically, we observed a \revision{convex quadratic} theta effect, a finding consistent with several previous lab studies using difficulty-manipulated paradigms \cite{Ewing2016, Fairclough2013, Berta2013, Chanel2011, Katahira2018} that most likely indicates task engagement levels. 
\revision{It should be noted, that our data are not sufficiently clear on whether the observed pattern shows a full inverted-U shape (e.g., as in \cite{Ewing2016, Fairclough2013}). While this is indicated by the fitted curve, the two-lines test only confirms a large slope change at the inflection point, but no significant negative slope after it. Thus, it is also possible that flow is not reduced much at very high theta activation (e.g., as in \cite{Katahira2018, Berta2013}). Together with the related work, it is most likely that the negative effect in the higher theta levels with flow is weaker than the positive effect between flow and theta in the lower theta levels. However, given the limited number of data points in our data subset after the inflection point, more research will be needed to confirm this.}
Nevertheless, the transfer of this observation into a real-world setting is an important contribution to the study of natural flow observation, as the finding emerges across various types of natural knowledge work. It might, therefore, be useful in the future for the development of flow experience classifiers that can infer flow without the need for additional behavioral data. However, as brain data is known to show high day-to-day variance \cite{Putze2022, vortmann2021}, we believe that longer study periods (possibly several weeks) will be needed to ascertain the efficacy and reliability of such models.
Achieving such study designs, might, however, be difficult with a gelled ear-EEG system. We used the open-cEEGrid system with the main goal of collecting clean neural signals. However, it is rather unlikely that this system will be usable on a daily basis, which can be seen by the fact that we had the participants come to the lab to prepare their skin and attach the electrodes correctly. Therefore, it is unclear if our results would easily translate to wearable EEG systems that use dry sensors, which are known to collect much more noise \cite{niso2023wireless}. However, as we find the present results encouraging, we believe it should be a next step to try and use gel-free ear-EEG systems like headphone EEG systems \cite{An2021, kartali2019real, Knierim2023} with similar and prolonged study designs. \\

Additionally, our study presents a novel finding: an inverted-U-shaped relationship between beta asymmetry and flow, which emerged exclusively in the natural work setting. Interestingly, we find this feature to be associated with both sub-dimensions of the flow scale: absorption and fluency, and without relationships to prominent flow confounds like arousal, valence, or mental workload. This indicates a flow-specific relationship.
Relationships between asymmetry features have previously only emerged in flow studies using mastery paradigms (where repeated task execution focused on fostering expertise and focus), which is arguably closer to the herein created natural work situation. 
Previous work attributed asymmetries during flow to motivational processes (frontal alpha - \cite{Labonte2016}) or verbal-analytic interference for motor tasks (temporal theta - \cite{Wolf2014, Kramer2007, DeKock2014}).
\revision{However, no research has so far reported on temporal beta asymmetries during flow. Therefore, we can only propose new possible explanations.
In neuroscientific research, parietal beta asymmetries have been linked to approach-avoidance motivation \cite{Schutter2001} - similar as frontal alpha asymmetry \cite{Metzen2021}. Thus, our quadratic temporal beta asymmetry finding might represent a neutral motivational state during flow (neither particularly avoiding nor approaching the current situation). However, as this would contrast the previous finding by \cite{Labonte2016} and does not integrate well with the common conception that flow is a state of high intrinsic motivation (i.e., approach motivation) \cite{Peifer2021}, this explanation does not seem most likely at the moment.} \\

\revision{Other studies have suggested that parietal beta asymmetries reflect attentional shifting, indicating varying levels of sustained attention \cite{Hale2014}. Thus, our quadratic temporal beta asymmetry finding may represent a balance between attentional flexibility and sustained attention during flow. This aligns with neurocognitive models of flow \cite{Harris2017} that highlight both the need for keeping focus on a task, but also being able to integrate new information (e.g., when flow is experienced during musical improvisation, both focus and flexibility are needed). However, as frontal brain regions are heavily involved in regulating such processes (also seen in recent flow EEG research - \cite{Tan2024}), additional studies with more comprehensive EEG setups will be needed to confirm this hypothesis.}
\revision{In a similar direction, our beta asymmetry finding 
may indicate elevated cognitive flexibility, the ability to adapt to new tasks or situations \cite{Ionescu2012, Ocklenburg2012}.
Cognitive flexibility is required in complex tasks - which have been linked to more interaction between the brain hemispheres \cite{Ocklenburg2012}. 
This could explain why the asymmetry was observed in complex tasks but not simpler ones like mental arithmetic.
A higher cognitive flexibility (represented by balanced beta activity), might be vital to efficiently bringing together the cognitive processes required to complete a complex task, facilitating the emergence of flow. 
However, it can at the moment not be excluded, that this finding might also be due to the nature of the tasks that might have required imbalanced involvement of creative, analytical, verbal, motor, and spatial or holistic processes. Therefore, future work would have to test this hypothesis using more controlled, albeit complex tasks.}

\section{Conclusion}
In this study, we explored the potential of monitoring flow experiences in everyday life utilizing a discreet, wearable EEG system positioned around the ear (open-cEEGrids). Our approach involved a descriptive analysis of flow in two natural work settings (academic thesis or software engineering projects) and a comparative analysis to a classic, difficulty-manipulated task. 
This study design allowed us to explore the feasibility of monitoring flow using brain data in natural settings, as we found that high flow intensities could be elicited in the knowledge work tasks. In addition, the combination of the two task paradigms allowed us to replicate and transfer a well-known EEG findings from the lab (a convex quadratic relationship between theta frequency power and flow) to the field.
These findings provide a valuable foundation for researchers for developing more robust predictive models of flow using physiological sensors, enhancing our understanding of how flow can be detected and measured in everyday life. \\

\revision{In addition, the results of this study also have important implications for future HCI research and technology development. Given the wearable EEG system's placement around the ear, it offers the advantage of being discreet and providing continuous, real-time data with high temporal precision. This makes it particularly suitable for integration into devices that are already part of daily life, such as sensorized headphones~\cite{Knierim2023, kartali2019real, Cherep2022}. Thus, future research could explore the application of flow monitoring in environments where headphones are naturally used, such as e-sports, virtual meetings, or media (e.g., music or video) production.
Here, real-time flow detection could inform dynamic interventions, such as recommending timely breaks when flow is low, suggesting adjustments to task difficulty or goals to maintain optimal engagement~\cite{Barros2018, Bartholomeyczik2024}, and to balance productivity and emotional well-being. Furthermore, future HCI research could explore how social interactions could be enriched through flow-related neurofeedback to foster shared flow experiences (e.g., in cooperative games~\cite{Labonte2016}), which have been considered as particularly rewarding and even social tie strengthening~\cite{Zumeta2016}.
Altogether, our findings thus provide ample opportunity to further study the development of adaptive technologies that foster flow and enhance both personal and social experiences.}

\begin{acks}
Funded by the German Research Foundation (Deutsche Forschungsgemeinschaft - DFG) – GRK2739/1 – Project Nr. 447089431 – Research Training Group: KD²School – Designing Adaptive Systems for Economic Decisions.
\end{acks}

\bibliographystyle{ACM-Reference-Format}
\bibliography{sample-base}
\end{document}